\documentclass[useAMS,usenatbib]{mn2e}

\usepackage{times,graphicx,amssymb,natbib,color}

\title[Convergence of self-gravitating disc simulations]{Convergence of simulations of self-gravitating accretion discs II: \\
Sensitivity to the implementation of radiative cooling and artificial viscosity} 
\author[W.K.M. Rice, S.-J. Paardekooper, D.H, Forgan and P.J. Armitage]{W.K.M. Rice$^{1}$\thanks{E-mail:
wkmr@roe.ac.uk}, S.-J. Paardekooper$^{2}$, D.H. Forgan$^{1}$ and P.J. Armitage$^{3,4}$\\
$^{1}$Scottish Universities Physics Alliance (SUPA), Institute for Astronomy, University of Edinburgh, Blackford Hill, Edinburgh, EH9 3HJ, UK \\
$^{2}$DAMTP, University of Cambridge, Wilberforce Road, Cambridge, CB30WA \\
$^{3}$JILA, 440 UCB, University of Colorado, Boulder, CO 80309-0440, USA \\
$^{4}$Department of Astrophysical and Planetary Sciences, University of Colorado, Boulder, USA}
\begin{document}

\date{Accepted 0000}

\pagerange{\pageref{firstpage}--\pageref{lastpage}} \pubyear{0000}

\maketitle

\label{firstpage}

\begin{abstract}
Recently it has been suggested that the fragmentation boundary in Smoothed Particle Hydrodynamic (SPH) and {\sc fargo} simulations of self-gravitating 
accretion discs with $\beta$ cooling do not converge as resolution is increased.  Furthermore, this recent work suggests that by carefully optimising the
artificial viscosity parameters in these codes it can be shown that fragmentation may occur for much longer cooling times than earlier work
suggests.  If correct, this result is intriguing as it suggests that gas giant planets could form, via direct gravitational collapse, reasonably
close to their parent stars.  This result is, however, slightly 
surprising and there have been a number of recent studies suggesting that the result is likely an indication of a numerical problem with the simulations.  One suggestion,
in particular, is that the SPH results are influenced by the manner in which the cooling is implemented.  We extend this work here and show that
if the cooling is implemented in a manner that removes a known numerical artefact in the shock regions, the fragmentation boundary converges to
a value consistent with earlier work and that fragmentation is unlikely for the long cooling times suggested by this recent work.  We also investigate the optimisation of the artificial
viscosity parameters and show that the values that appear optimal are likely introducing numerical problems in both the SPH and {\sc fargo} simulations.
We therefore conclude that earlier predictions for the cooling times required for fragmentation are likely correct and that, as suggested by this earlier 
work, fragmentation cannot occur in the inner parts ($r < 50$ au) of typical protostellar discs.
\end{abstract}


\section{Introduction}

\noindent
If a disc around a central object is sufficiently massive, its own self-gravity may play an
important role in its evolution through the growth of the gravitational instability \citep{safronov60,goldreich65}.  An infinitesimally thin disc
is susceptible to the growth of an axisymmetric gravitational instability if the $Q$ parameter \citep{toomre64}
\begin{equation}
Q = \frac{c_s \kappa}{\pi G \Sigma} < 1,
\label{eq:Q}
\end{equation}
where $c_s$ is the sound speed in the disc, $\kappa$ is the local epicyclic frequency (equal
to the angular velocity, $\Omega$, in a Keplerian disc), $G$ is the gravitational constant,
and $\Sigma$ is the disc surface density.  Global, non-axisymmetric perturbations can, however, grow
for $Q$ values greater than $1$, with simulations suggesting that the stability criteria in
global discs is $Q < 1.5 - 1.7$ \citep{durisen07}. 

It is now, however, quite well understood that the $Q$ parameter alone does not determine
the ultimate evolution of a self-gravitating accretion disc.  The evolution is determined
by both the value of $Q$ and by the rate at which the disc is able to lose energy \citep{pickett98,gammie01}.
The current picture is that for long cooling times, the disc will settle into a state of marginal
stability \citep{paczynski78} in which the instability acts to transport angular momentum 
outwards, allowing mass to accrete onto the central object \citep{lin87,laughlin94,lodato04,mejia05}. For
short cooling times, however, the disc may become sufficiently unstable to fragment
and form bound objects \citep{kuiper51}.  This has been suggested as a mechanism for forming
gas giant planets in discs around young stars \citep{boss98, boss00} or stars in discs around
supermassive black holes \citep{shlosman89,goodman03,bonnell08}.

Two-dimensional, shearing sheet simulations \citep{gammie01}, using a specific heat ratio of $\gamma = 2$, indicated that the boundary between
fragmentation and a quasi-steady, self-gravitating state occurred at a cooling time of
\begin{equation}
\tau_{\rm c} = 3 \Omega^{-1}.
\label{eq:tcool_fragbound}
\end{equation}
\citet{rice03} found a similar result using three-dimensional Smoothed Particle Hydrodynamics (SPH)
simulations. As already mentioned, in a quasi-steady state the gravitational instability acts to transport 
angular momentum outwards. In many instances it is appropriate to assume that angular momentum
transport is driven by disc viscosity which \citet{shakura73} suggest has the form
\begin{equation}
\nu = \alpha c_s H,
\label{eq:alpha_visc}
\end{equation} 
where $H=c_s/\Omega$ is the disc scaleheight. This form, however, assumes that the viscosity depends only
on local disc properties.  Given that disc self-gravity is inherently global, such a form is not
necessarily suitable for characterising angular transport in self-gravitating discs.  However,
if $Q \sim 1$ and if the disc mass is less than about half that of the central star, a local approximation
appears to be a suitable representation \citep{balbus99, lodato04, lodato05, forgan11}.  

Since a quasi-steady state is one in which the cooling is balanced by an effective viscous heating, one
can relate the viscosity to the cooling time through \citep{pringle81,gammie01}
\begin{equation}
\alpha=\frac{4}{9 \gamma (\gamma - 1) \tau_{\rm c} \Omega},
\label{eq:alpha_tcool}
\end{equation}
where $\gamma$ is the specific heat ratio.  Using different values of $\gamma$, \citet{rice05}
showed that rather than the fragmentation boundary depending on the cooling time, $\tau_{\rm c}$,
it depends on the stresses in the disc, as represented by $\alpha$.  In agreement with \citet{gammie01}
their results indicate that self-gravitating discs can maintain a quasi-steady state if $\alpha < 0.06$
and will fragment if the required stress exceeds $\alpha > 0.06$.  

\citet{cossins09} used analytic calculations and three-dimensional numerical simulations to investigate further the energy
balance in self-gravitating discs and found that the perturbation amplitude, $\delta \Sigma/\Sigma$, 
is related to the cooling time through
\begin{equation}
\frac{\delta \Sigma}{\Sigma} = \frac{1}{\sqrt{\beta_{\rm cool}}},
\label{eq:perturb}
\end{equation}
where $\beta_{\rm cool} = \tau_c \Omega$. Using two-dimensional shearing-sheet simulations, \citet{rice11} showed, 
similarly, that $\delta \Sigma/\Sigma \propto \alpha$.
The basic picture that has therefore been developed is that a self-gravitating accretion disc will settle into a quasi-steady state
in which cooling is balanced by heating driven by the gravitational instability.  In such a state, the perturbation
amplitudes will depend on the cooling rate (or, equivalently, on the level of stress in the disc) and if these perturbations
are sufficiently large, they become non-linear, the disc is unable to maintain a quasi-steady state and instead
fragments into bound objects.  What makes this general picture attractive is that there is reasonable agreement across
a wide-range of different types of simulations including two-dimensional shearing sheet simulations \citep{gammie01,rice11},
three-dimensional grid-based simulations \citep{mejia05,boley07,steiman-cameron13}, and 
three-dimensional SPH simulations \citep{rice05,cossins09}. 

Recent work \citep{meru11} has, however, shown that three-dimensional SPH simulations that fix $\beta_{\rm cool}$, 
do not converge to a well-defined
fragmentation boundary as resolution is increased.  Their highest resolution simulations suggested that fragmetation
could occur for cooling time $\tau_{\rm c} > 10 \Omega^{-1}$.  Given that the Jeans mass of a typical fragment is well-resolved
in an SPH simulation even for quite modest resolutions \citep{bate97,rice12}, this result is quite surprising.  There have been a number
of attempts to understand this result.  \citet{lodato11} suggest that numerical viscosity may influence disc thermodynamics
more than originally thought and hence that simulations may require higher resolutions than indicated by earlier
calculations.  \citet{paardekooper11} use two-dimensional grid-based simulations to show that the lack of convergence
could be related to edge effects in simulations with very smooth initial conditions.  There is also some
suggestion that disc fragmentation may have a stochastic nature \citep{paardekooper12}.  

\citet{michael12} considered convergence in three-dimensional, grid-based, self-gravitating disc simulations.  This work, 
however, didn't directly address
convergence of the fragmentation boundary, but instead considered convergence of the properties of a quasi-steady, self-gravitating
disc.  Their results were consistent with fragmentation requiring $\alpha > 0.06$, but couldn't really make any strong statements
about convergence of the fragmentation boundary. \citet{steiman-cameron13}, extended the work of \citet{michael12} to consider
simulations with radiative cooling and found that the properties of these discs did not converge in the outer, optically-thin 
regions.  One of their conclusions was that there may be issues with convergence in regions where the optical depth is of order 
unity.

It has, however, been suggested \citep{rice12} that the lack of convergence of
the fragmentation boundary in the \citet{meru11} simulations was simply a consequence of the manner in which the cooling was implemented. 
\citet{rice12} suggest that it may be related to the known problem - in many SPH implementations - 
of an unphysical discontinuity in the thermal energy (pressure)
at contact discontinuities \citep{price08, price12} which, if not corrected for, could lead to regions with enhanced cooling.
\citet{rice12} suggested that this could be solved by using a form of the cooling that smooths across each SPH particle's 
neighbour sphere.  In their simulations, fragmentation occurred only for cooling times $\tau_{\rm c} < 9 \Omega^{-1}$ but they
could not claim convergence as their highest resolution simulations fragmented for a slightly longer cooling time than the
value towards which the others appeared to be converging.

\citet{meru12} have recently extended this convergence work to consider how it is affected by artificial viscosity in both
SPH simulations and in {\sc fargo} grid-based simulations. They consider various artificial viscosity parameters and settle on the values that maximise
the value of $\beta_{\rm cool}$ for which fragmentation can occur.  In a quasi-steady state the disc is in thermal equilibrium with the imposed
cooling balanced by heating from both the instability and from artificial viscosity.  Ideally, the artificial heating should be 
minimised.  The perturbation amplitudes should depend on the strength of the instability \citep{cossins09,rice11},
therefore choosing artificial viscosity parameters that maximise the value of $\beta_{\rm cool}$ at which fragmentation can occur
should minimise the level of artificial heating.

There are, however, a few issues with this that will be discussed in more detail in a later section.  Artificial
viscosity is typically introduced so as to resolve shocks.  In SPH, however, the artificial viscosity
operates even in the absence of shocks, producing an artificial dissipation that should, ideally, be minimised.
The way in which the instability heats the disc is through dissipation at shocks and so varying the viscosity
parameters can reduce the artificial dissipation, but can also have an impact on the shock heating.  
Given that \citet{rice12} suggest that the lack of convergence
seen in \citet{meru11} could be due to an unphysical structure in the shock regions, changing the shock structure
could exacerbate this problem.  Additionally, in {\sc fargo}, the artificial viscosity only operates at shocks and doesn't produce 
any kinematic viscosity, so shouldn't introduce any artificial heating. Any artificial diffusion in {\sc fargo} should
occur at the grid scale and therefore shouldn't depend on the artificial viscosity parameters.  

In this paper we extend the work of \citet{rice12} to show that their suggested cooling formalism does indeed
appear to lead to convergence of the fragmentation boundary and that, consistent with earlier work, fragmentation
requires $\beta_{\rm cool} < 8$ for $\gamma = 5/3$.  We then extend this to consider how this results depends on
the artificial viscosity in SPH to establish if the values suggested by \citet{meru12} are indeed optimal. We also discuss 
their results obtained using {\sc fargo}. In Section 2 
we briefly describe Smoothed Particle Hydrodynamics (SPH).  In Section 3 we consider the SPH results unsing 
the cooling formalism suggested by \citet{rice12}.  In Section 4 we address the results \citet{meru12} obtained
using {\sc fargo} and in Section 5 we discuss these results and conclude.

\section{SMOOTHED PARTICLE HYDRODYNAMICS}

\subsection{The basic formalism}
SPH is a Lagrangian hydrodynamic formalism in which a fluid, or gas, is represented by pseudo-particles 
(see e.g., \citet{benz90, hernquist89, monaghan92}).  Each particle is assigned a a mass ($m$), position
($x,y,z$), velocity ($v_x, v_y, v_z$) and internal energy per unit mass ($u$).  There are many descriptions
of SPH (e.g., \citealt{benz90, monaghan92}), so we won't repeat the details here.
Basically, fluid/gas density is calculated via interpolation across the mass distribution.  Pressure is determined
via an equation of state. Gravitational forces can either be calculated by direct summation, or - more
commonly - using a TREE code \citep{barnes86}.  The momentum and energy equations are in a form suitable for this
Langrangian formalism and so the particles velocities are updated using the gravitational and pressure forces on each particle, 
the positions are updated using the velocity of each particle, and the internal energy changes via $PdV$ work, viscous dissipation 
and cooling.

\subsection{Introducing cooling}
To investigate the evolution of self-gravitating discs, a cooling time, $\tau_c$, of the 
of the following form is typically used \citep{gammie01,rice03}:
\begin{equation}
\tau_c = \beta_{\rm cool} \Omega^{-1}.
\label{eq:tcool}
\end{equation}
This is typically added to the energy equation by assuming that the thermal energy of each particle, $u_j$, decays
with an e-folding time given by $\tau_c$.  Hence, the energy equation becomes,
\begin{equation}
\frac{d u_j}{d t} = \frac{1}{2} \sum_i \left( \frac{P_i}{\rho_i^2} + \frac{P_j}{\rho_j^2} \right) {\bf v}_{ij} \cdot 
\nabla_i W \left( | {\bf r}_i - {\bf r}_j |; h \right) - \frac{u_j}{\tau_c}, 
\label{eq:energ_eq}
\end{equation}
where the sum is over all the neighbours, $i$, of particle $j$, ${\bf v}_{ij}$ is the velocity difference
between particle $i$ and particle $j$, $W$ is the smoothing kernel used to interpolate across the 
neighbours of particle $j$, and $h$ is the smoothing length that defines the volume of the neighbour sphere. 
As already mentioned, the unsmoothed - or individual particle - internal energy has an unphysical discontinuity
at the contact discontinuity behind shock waves \citep{price08, price12}. Consequently, \citet{rice12}
suggested that the cooling be distributed across the neighbour sphere.  Their suggestion was that the standard 
cooling term
implementation
\begin{equation}
\left( \frac{d u_j}{dt} \right)_{\rm cool} = -\frac{u_j}{\tau_c},  
\label{eq:basic_cool}
\end{equation}
should be replaced with
\begin{eqnarray}
\left( \frac{d u_j}{d t} \right)_{\rm cool} &=& -\frac{1}{\tau_c} \frac{W(0,h_j) m_j u_j}{\rho_j} \nonumber \\
\left( \frac{d u_{i \ne j}}{d t} \right)_{\rm cool} &=& - \frac{1}{\tau_c} \frac{m_i}{\rho_i} u_i W(|\textbf{r}_j-\textbf{r}_i|,h).
\label{eq:SmoothCool2}
\end{eqnarray}
Their suggestion was that the upper of the two equations in Equation (\ref{eq:SmoothCool2}) would be applied to particle
$j$, while the lower of the two equations would be applied to the neighbours $i$ of particle $j$.  This would ensure that
the cooling rate associated with particle $j$ would be
\begin{equation}
\left( \frac{d u_j}{d t} \right)_{\rm cool} = -\frac{1}{\tau_c}\sum_i \frac{m_i}{\rho_i} u_i W(|\textbf{r}_j-\textbf{r}_i|,h).
\label{eq:totalcoolj}
\end{equation}
In regions without discontinuities this will be the same as the cooling rate given by Equation (\ref{eq:basic_cool}) \citep{rice11}. 
At the discontinuities in the shock regions, Equation (\ref{eq:totalcoolj}) will use the physically correct interpolated
value for the internal energy and will ensure that the unphysical jump in thermal energy at the contact discontinuity
does not artificially enhance the cooling in that region. 

The results presented in \citet{rice12} using the cooling form shown in Equation (\ref{eq:SmoothCool2}) suggested that
the simulations were converging towards a fragmentation boundary between $\beta_{\rm cool} = 6$ and $\beta_{\rm cool} = 7$. Their highest
resolution simulation (10 million particles), however, fragmented between $\beta_{\rm cool} = 8$ and $\beta_{\rm cool} = 9$ and hence they could not claim convergence.  Their
highest resolution simulation was, however, only a ring of 4 million particles that would have had the same resolution
as a full 10 million particle simulation.  Strictly speaking, it wasn't exactly the same conditions as the other lower-resolution
simulations. Here, we present results from a single 10 million particle run using the cooling form presented in
Equation (\ref{eq:SmoothCool2}) and that indicates that a stronger constraint on the convergence of the fragmentation boundary.

\subsection{Artificial viscosity}
In SPH, the artificial viscosity has two main roles; to prevent particle interpenetration and to resolve shock waves. 
The viscosity can consequently be thought of as having a shear component and a bulk component \citep{monaghan85}, with
the bulk component acting very like a Von Neumann-Richtmeyer viscosity used to resolve shocks in many grid-based codes.
A way in which to introduce viscosity in SPH, and what is used in all the simulations presented here, 
is to use the following to determine the viscosity term between
particles $i$ and $j$;
\begin{equation}
  \Pi_{ij}=\left\{
   \begin{array}{cc}
      \frac{-\alpha_{\rm SPH} \overline{c}_{ij} \mu_{ij} + \beta_{\rm SPH} \mu_{ij}^2}{\overline{\rho}_{ij}} &  {\bf v}_{ij} \cdot {\bf r}_{ij} < 0;  \\
      0 & {\bf v}_{ij} \cdot {\bf r}_{ij} > 0.
   \end{array}\right. 
\label{eq:visc}
\end{equation}
The terms $\overline{c}_{ij}$ and $\overline{\rho}_{ij}$ are the average sound speed and density for particles $i$ and $j$.
The terms ${\bf v}_{ij}$ and ${\bf r}_{ij}$ are $({\bf v}_i - {\bf v}_j)$ and $({\bf r}_i - {\bf r}_j)$ respectively. The 
coefficients $\alpha_{\rm SPH}$ and $\beta_{\rm SPH}$ determine the strength of the two viscosity terms and the term $\mu_{ij}$ is given
by
\begin{equation}
\mu_{ij} = \frac{h {\bf v}_{ij} \cdot {\bf r}_{ij}}{{\bf r}_{ij}^2 + \eta^2},
\label{eq:mu}
\end{equation}
where $\eta$ is a softening term that prevents the denominator in Equation (\ref{eq:mu}) from
ever being zero. It should be clear that Equation (\ref{eq:visc}) only operates when particles 
are converging and hence will prevent interpenetration and resolve shocks. However, this form
of the viscosity does not distinguish between converging flows and shear flows and hence 
this viscosity can also transport angular momentum and can, therefore, heat the system in the absence of shocks \citep{cartwright10}.

The viscosity coefficients typically satisfy $\beta_{\rm SPH} = 2 \alpha_{\rm SPH}$ and commonly 
used values, in self-gravitating disc simulations, are $\alpha_{\rm SPH} = 0.1$ and $\beta_{\rm SPH} = 0.2$.  
One reason for using $\beta_{\rm SPH} = 2 \alpha_{\rm SPH}$ is that it ensures that the first term in
Equation (\ref{eq:visc}) dominates when the convergence is slow, while the second term dominates when
convergence is rapid.  The $\beta_{\rm SPH}$ term is essentially necessary so as to handle high Mach-number shocks
\citep{monaghan92}. Consequently, most studies \citep{murray96,lodato04} only consider
the $\alpha_{\rm SPH}$ term when determining the dissipation due to artificial viscosity in shear flows. Optimally
the $\alpha_{\rm SPH}$ value should be set so as to minimise artificial dissipation while still preventing
particle interpenetration.

\citet{meru12} point out that the dissipation associated with the $\beta_{\rm SPH}$ term is not actually
negligible but is about a factor of $2$ smaller than that associated with the $\alpha_{\rm SPH}$ term when
$\alpha_{\rm SPH} = 0.1$ and $\beta_{\rm SPH} = 0.2$. They, therefore, conclude that one should optimise in terms of
both $\alpha_{\rm SPH}$ and $\beta_{\rm SPH}$ and conclude that the optimal values are 
$\alpha_{\rm SPH} = 0.1$, $\beta_{\rm SPH} = 2$.  This
was largely based on simulations that maximised the value of $\beta_{\rm cool}$ - which determines
the cooling time - for which fragmentation could
occur.  Maximising the cooling time at which fragmentation occurs, suggests that one has minimised the amount of artificial
dissipation and so is an attractive strategy to adopt.  However, as we'll discuss in more detail later,
this may not necessarily be the case and so, here, we investigate how these values of $\alpha_{\rm SPH}$
and $\beta_{\rm SPH}$ influence the fragmentation boundary when using the modified cooling form
proposed by \citet{rice12}.  

\subsection{Simulation setup}
All of the SPH simulations presented here have the same basic setup as those presented by \citet{meru12}.
They have a central star with mass $M_* = 1$ surrounded by a disc extending from $r_{\rm in} = 0.25$ to
$r_{\rm out} = 25$, with a mass of $M_{\rm disc} = 0.1 M_*$, an initial surface density profile
of $\Sigma \propto r^{-1}$, and with an initial minimum $Q$ parameter of $Q$ = 2. 
We impose a cooling of the form described by Equation (\ref{eq:tcool}),
but that is either implemented as in \citet{meru12} - which we call basic cooling - or in the modified
manner suggested by \citet{rice12} - which we call smoothed cooling.  We consider various resolutions,
ranging from 250000 particles to 10 million particles. In all our simulations we take $\alpha_{\rm SPH} = 0.1$,
but consider both $\beta_{\rm SPH} = 0.2$ and $\beta_{\rm SPH} = 2$.  

\section{Results}
\label{sec:results}
\subsection{Convergence using smoothed cooling}
In the work of \citet{rice12} they considered full simulations using 250000, 500000, and 2 million particles,
but represented a 10 million particle simulation using a simulation with 4 million particles with a mass
of $M_{\rm disc} = 0.04 M_*$ and that extended from $r_{\rm in} = 15$ to $r_{\rm out} = 25$.  These
parameters were chosen so as to have the same properties, in that region, as a full 10 million particle simulation.   
In \citet{rice12} the 500000 and 2 million particle simulations fragmented at between $\beta_{\rm cool} = 6$ 
and $\beta_{\rm cool} = 7$ and appeared to be converging.  The pseudo-10 million particle simulation, however, fragmented between
$\beta_{\rm cool} = 8$ and $\beta_{\rm cool} = 9$ and hence they could not claim convergence.

\begin{figure}
\begin{center}
\includegraphics[scale = 0.2]{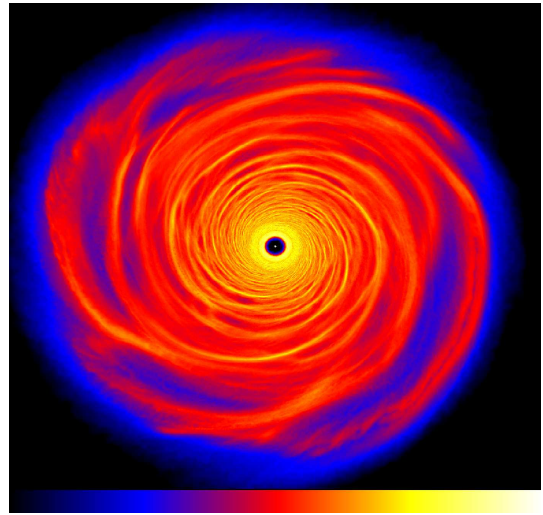}
\caption{Final state of a full 10 million particle simulation using smoothed cooling with $\beta_{\rm cool} = 8$.
At this stage (after 6.5 outer rotation periods) the disc has settled into a quasi-steady state and there is no evidence
of fragmentation.}
\label{fig:10million}
\end{center}
\end{figure}

We have since managed to complete a full 10 million particle simulation with $\beta_{\rm cool} = 8$ which, after
6.5 outer rotation periods, shows no signs of fragmentation. The state of the simulation at this time
(5120 code units or 815 orbits at $r = 1$) is shown in Fig. \ref{fig:10million}. There is clearly 
lots of spiral structure, but no evidence of fragmentation.  We also include a table with the 
results from the simulations of \citet{rice12} together with this new result using 10 million particles.   

\begin{table}
\centering
\begin{minipage}{140mm}
  \caption{List of the simulations using smoothed cooling.\label{tab:sims}}
  \begin{tabular}{c || ccc}
  \hline
  \hline
   Simulation & No. of particles &  $\beta_{\rm cool}$ & Fragment?  \\
 \hline
  1 & 250000 & 4 & Yes  \\
  2 & 250000 & 4.5 & Yes \\
  3 & 250000 & 5 & No  \\
  4 & 250000 & 6 & No  \\
  5 & 250000 & 7 & No  \\
  6 & 500000 & 5 & Yes  \\
  7 & 500000 & 6 & Yes \\
  8 & 500000 & 7 & No  \\
  9 & 500000 & 8 & No \\
  10 & 2000000 & 5 & Yes \\
  11 & 2000000 & 6 & Yes \\
  12 & 2000000 & 7 & No \\
  13 & 2000000 & 8 & No \\
  14 & 10000000 & 8 & No \\
 \hline
  \hline
\end{tabular}
\end{minipage}
\end{table}

\begin{figure}
\begin{center}
\includegraphics[scale=0.5]{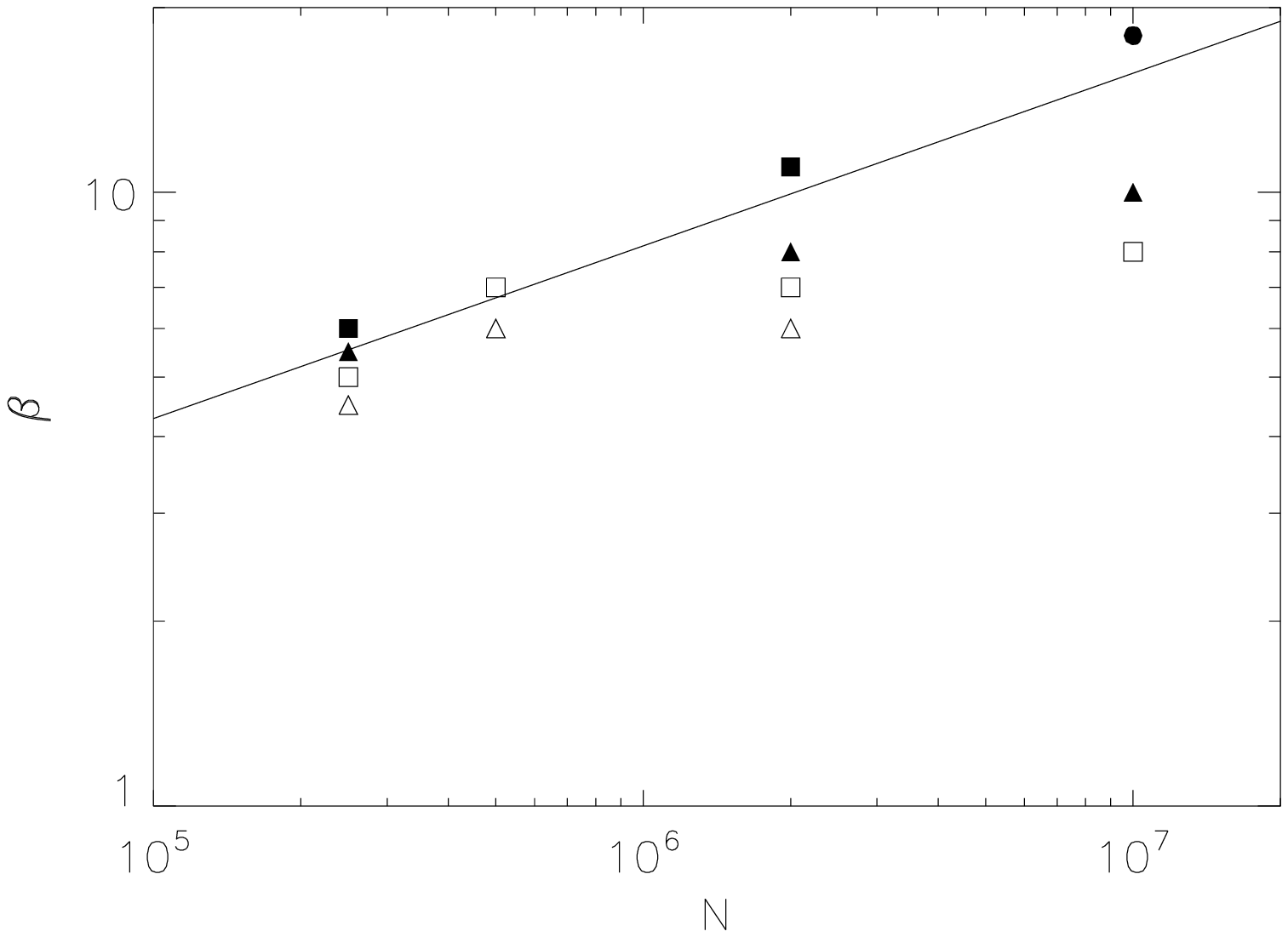}
\caption{Fragmentation boundary from \citet{meru11} (filled symbols) together with the results 
from \citet{rice12} and the new 10 million particle result from this work (open symbols).  The
triangles are for the largest value of $\beta_{\rm cool}$ for which fragmentation occured.  The
squares are for the smallest value of $\beta_{\rm cool}$ for which a quasi-steady state is
reached.  The circle is for one simulation from \citet{meru11} which was regarded as borderline.
The line is from \citet{meru11} and illustrates the lack of convergence seen in their simulations.
The results from \citet{rice12} together with the new result from this work would indicate that
the fragmentation boundary has converged to a value of $\beta_{\rm crit} < 8$.}
\label{fig:fragbound}
\end{center}
\end{figure}

Figure \ref{fig:fragbound} is an updated version of that presented by \citet{rice12}.  It shows $\beta_{\rm cool}$ plotted
against particle number.  The squares are for those simulations with the largest values of $\beta_{\rm cool}$ that fragmented 
and in which the fragments survived and became sufficiently dense that the simulation effectively stopped.  
The triangles are for those with the lowest values of $\beta_{\rm cool}$ that did not fragment.  The single filled circle is 
for a simulation that was regarded as borderline \citep{meru11}.  The data points are from \citet{rice12}
(open symbols and particle numbers of 250000, 500000, and 2million), \citet{meru11} (filled symbols),
and this work (10 million particle simulation).  

The line in Figure \ref{fig:fragbound} is the same as that included by \citet{meru11} to show that
their simulations do not converge to a well-defined value of $\beta_{\rm crit}$ that defines the fragmentation boundary.
In their later work \citep{meru12}, however, they carry out many more simulations and suggest that the fragmentation
boundary is actually converging towards $\beta_{\rm crit} \sim 17.4$. However, even their highest resolution simulations (16 million
particles) haven't technically converged as they fragment between $\beta_{\rm cool} = 10$ and $\beta_{\rm cool} = 12$. Figure \ref{fig:fragbound}
shows that using smoothed cooling (which should only differ from basic cooling in the shock regions) results in
fragmentation requiring $\beta_{\rm cool} < 8$ for all resolutions considered. To have a better sense of whether these have converged or not, we should probably
do a 10 million particle simulation with $\beta_{\rm cool} = 7$, but this would take a significant amount of time and it already
seems clear that smoothed cooling results in fragmentation requiring $\beta_{\rm cool} < 8$ for all particle numbers considered and
that the converged value would be between $\beta_{\rm cool} = 6$ and $\beta_{\rm cool} = 8$.

Figure \ref{fig:fragbound} also suggests that simulations with 500000 particles are close to being converged, if not
actually converged. This is numerically sensible given that with $\alpha_{\rm SPH} = 0.1$ and $\beta_{\rm SPH} = 0.2$,
artificial dissipation should be providing less than 10\% of the heating for $\beta_{\rm cool} < 10$ \citep{lodato04}. Furthermore,
the outer half of such a disc resolves the Jeans Mass at $Q = 1$ \citep{rice12} and so should suitably resolve any possible fragmentation.
Given that the smoothed cooling formalism seems like a reasonable manner in which to impose the cooling and which should only
differ from basic cooling in the shock regions (where it will act to remove the unphysical discontinuity in thermal energy at the contact
discontinuity) one could conclude that fragmentation requires (for $\gamma = 5/3$) $\beta_{\rm cool} < 8$ and that the lack of convergence
seen in \citet{meru11} and \citet{meru12} is an entirely numerical artifact related to the manner in which they impose their cooling.

We should add that if our interpretation of the reason for this convergence issue has merit, it may be possible to continue
using the basic cooling formalism as long as some form of heat conduction is included in the SPH simulations 
so as to remove the non-physical jump in the thermal energy at contact discontinuities \citep{price08}.  We plan to investigate this possibility
in future work.

\subsection{The influence of artificial viscosity - basic cooling}
To investigate the lack of convergence of the fragmentation boundary in SPH simulations with $\beta$-cooling,
\citet{meru12} considered the influence of artificial viscosity.  As discussed earlier, the artificial
viscosity has two terms, one that largely acts to prevent particle interpenetration and one that acts to
resolve high Mach-number shocks.  In principle, the artificial viscosity should only act on converging flows.  However, 
in a disc simulation it is unable to distinguish between converging flows and shear flows and so it also produces
a shear viscosity that transports angular momentum and hence, even in the absence of shocks, produces dissipation.  
Ideally, the artificial viscosity parameters should be optimised so as to minimise the level of artificial dissipation and hence ensure 
that ``real" dissipation (such as that associated with the gravitational instability) dominates.

Previous work \citep{lodato04,rice05,cossins09} has used $\alpha_{\rm SPH} = 0.1$ and $\beta_{\rm SPH} = 0.2$.
\citet{meru12} vary $\alpha_{\rm SPH}$ and $\beta_{\rm SPH}$, in simulations with 250000 particles, to consider
how these parameters influence the fragmentation boundary.  The goal, in some sense, is to maximise $\beta_{\rm crit}$
(i.e., the critical value of $\beta_{\rm cool}$ at which fragmentation occurs) as this
would imply that this has minimised the level of artificial dissipation.  With fixed $\beta_{\rm SPH}$ they find
that $\beta_{\rm crit}$ is maximised for $\alpha_{\rm SPH} = 0.1$.  For fixed $\alpha_{\rm SPH}$ they claim that
their results suggests that the optimal value for $\beta_{\rm SPH}$ is $\beta_{\rm SPH} = 2$.  However, their own figure suggests that
$\beta_{\rm crit}$ is approximately constant for $\beta_{\rm SPH} < 0.2$, that it rises as $\beta_{\rm SPH}$ is 
increased from 0.2 to 2, and then become constant again for $\beta_{\rm SPH} > 2$. They haven't really maximised
$\beta_{\rm crit}$; they appear to have found a plateau.  This would suggest that any value of $\beta_{\rm SPH} > 2$ would be suitable, 
which is a little surprising as $\beta_{\rm SPH}$ essentially resolves shocks and determines the shock structure. 
Larger values will tend to result in broader shock fronts and so the norm would be to make $\beta_{\rm SPH}$ as small
as possible.  That their results suggest a step change between $\beta_{\rm SPH} = 0.2$ and $\beta_{\rm SPH} = 2$
might indicate a numerical issue, rather than an optimisation of $\beta_{\rm SPH}$.

\citet{meru12} then repeated their earlier simulations \citep{meru11} using $\alpha_{\rm SPH} = 0.1$
and $\beta_{\rm SPH} = 2$.  These simulations resulted in values for $\beta_{\rm crit}$ that were
typically at least 50\% greater than that obtained using $\alpha_{\rm SPH} = 0.1$ and $\beta_{\rm SPH} = 0.2$. Their analysis
suggested that these simulations were converging towards $\beta_{\rm crit} = 29.2$, considerably higher than
the value ($\beta_{\rm crit} = 17.4$) obtained \citep{meru11} with basic cooling using $\alpha_{\rm SPH} = 0.1$, $\beta_{\rm SPH} = 0.2$. This would
tend to imply that the larger value of $\beta_{\rm SPH}$ has resulted in a reduced level of artificial
dissipation.  

There are, however,  a number of issues with this interpretation.  For example, one would expect the level of artificial dissipation
to decrease as particle number increases.  The value of $\beta_{\rm crit}$ should therefore converge, for large
enough particle number, to a value independent of $\beta_{\rm SPH}$.  The \citet{meru12} analysis does not seem
to show any such convergence.  Additionally, if the correct value of $\beta_{\rm crit}$ is more than 50\% higher than suggested
by earlier SPH simulations this would imply that artificial dissipation provided at least one-third of the heating
in these earlier SPH simulations.  Earlier work \citep{lodato04} considered the level of artificial dissipation
and estimated, for simulations with 250000 particles, that it would provide less than 10\% of the dissipation
for $\beta < 10$.  

\citet{forgan11} used 500000 particle SPH simulations to investigate how self-gravitating
discs evolve in the presence of realistic cooling.  Their estimate of the level of artifical dissipation using
the $\alpha_{\rm SPH}$ term only, as suggested by \citet{murray96}, lead them to suggest that the artificial dissipation
would dominate inside 10 - 20 AU (consistent with a similar analysis by \citet{clarke09}).  This was consistent
with the results of their simulations, as the gravitational stresses very quickly became negligible inside
10 - 20 AU \citep{forgan11}.  The simulations by \citet{forgan11} are not, however, consistent with artificial dissipation being at least a factor of 3 higher than
expected as that would have lead to the artificial dissipation dominating inside 30 AU, rather than only inside 10 - 20 AU.

It has, however, been shown that there are circumstances in which $\beta_{\rm SPH}$ values of $2$ or greater may be 
necessary so as to reduce the level of unphysical, random particle motions \citep{price10}. However, this appears to
be mainly relevant for high Mach number flows ($M \sim 10$). The turbulent velocities in self-gravitating
disc simulations are typically subsonic \citep{forgan11b} and so it is likely that the linear SPH artificial viscosity 
term ($\alpha_{\rm SPH}$) will play the dominant role in reducing this particle noise. Furthermore, 
the only source of energy for random particle motions is the kinetic energy of the flow itself. Therefore
if such artificial turbulence was being generated and dissipated, this should transport angular momentum and be reflected in the calculated stresses.
As discussed above, however, previous work \citep{lodato04,forgan11} appears not to be consistent with the level of 
such turbulence being significantly greater than basic estimates suggest \citep{murray96}.

\subsection{The influence of artificial viscosity - smoothed cooling}
By varying the artifical viscosity parameters, \citet{meru12} have suggested that the fragmentation boundary, as parametrised by
$\beta_{\rm crit}$, is at least 50\% higher than previously thought.  As suggested above, however, this does appears to be inconsistent with
earlier work \citep{lodato04, forgan11}.  Their choice
of artificial viscosity parameters was, however, motivated by a sense that these parameters minimised the level of
artificial dissipation.  Consequently, if this is indeed the case, we would expect to see a similar result if we used these parameters together with
the smoothed cooling suggested by \citet{rice12} (see Equation (\ref{eq:SmoothCool2})).  

\begin{figure}
\begin{center}
\includegraphics[scale=0.2]{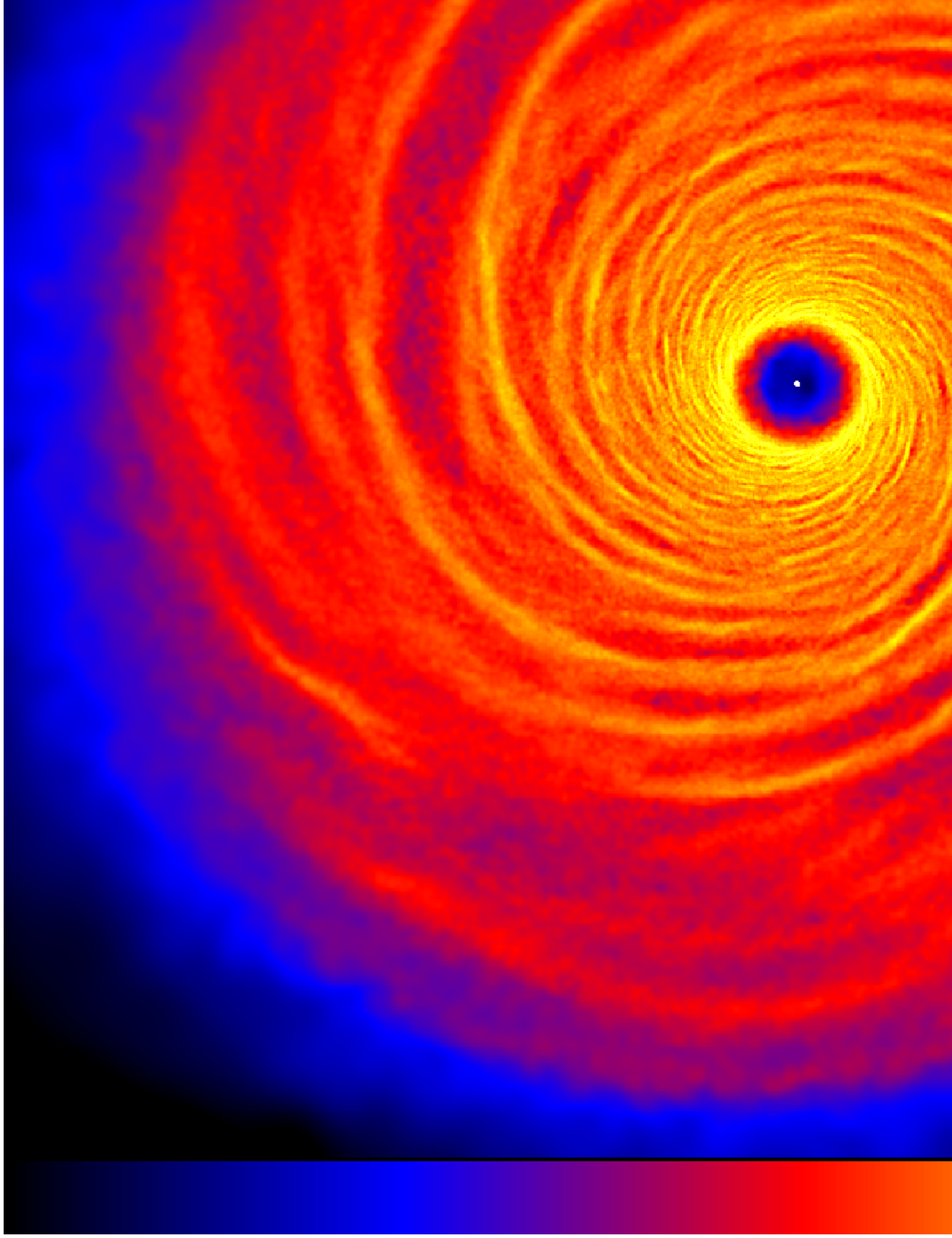}
\includegraphics[scale=0.2]{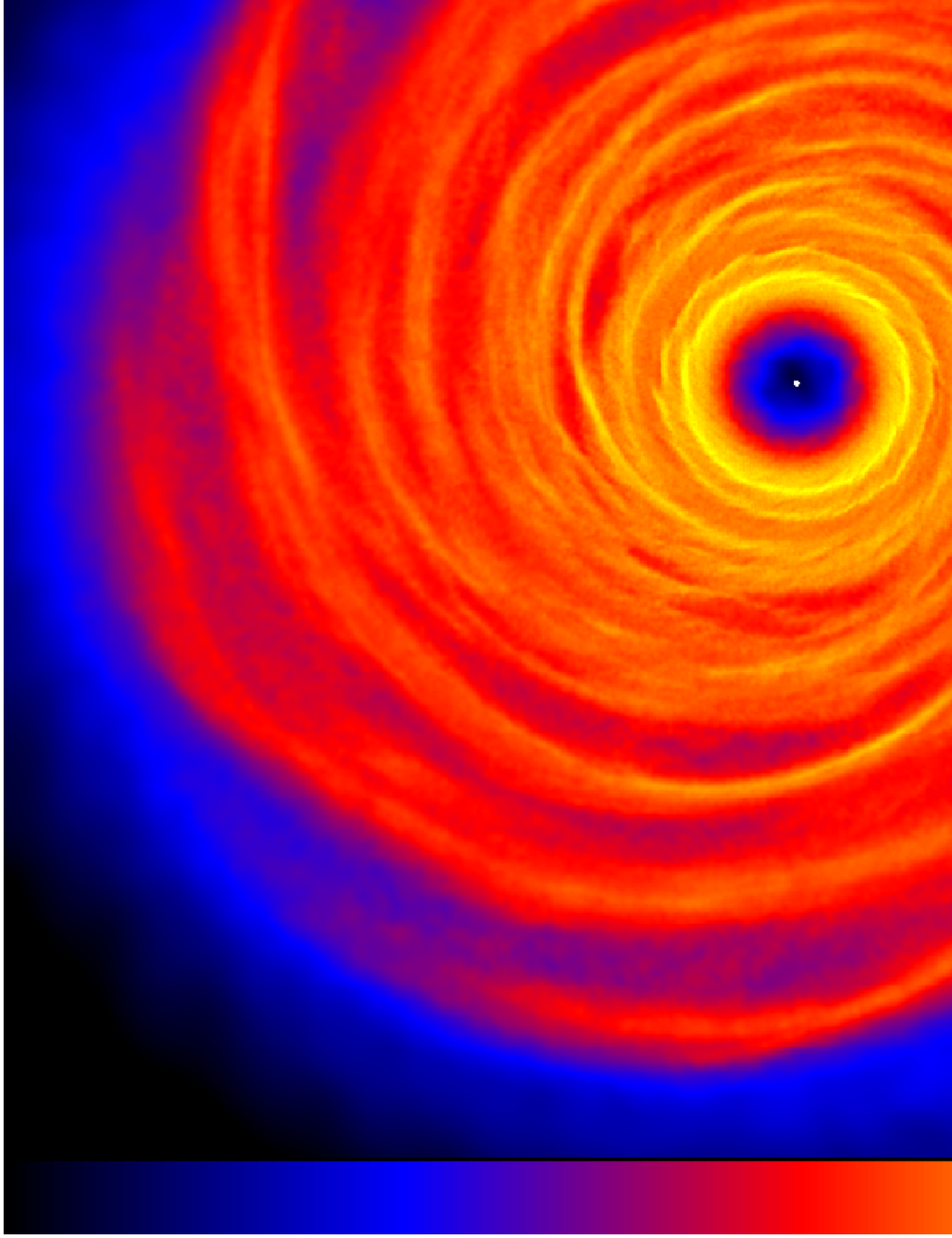}
\caption{The final states of two 500000 particles simulations both of which used $\beta_{\rm cool} = 8$ with smoothed cooling.
The top figure is from a simulation using $\beta_{\rm SPH} = 0.2$, while the bottom is from one that used $\beta_{\rm SPH} = 2$.  Although similar,
there are clear differences.  Using $\beta_{\rm SPH} = 2$ seems to reduce the level of noise in the outer
parts of the disc, but also smears out the spiral structure in the inner regions and increases the size of the inner hole.}
\label{fig:500k_comparison}
\end{center}
\end{figure}

To investigate the influence of artificial viscosity when using smoothed cooling, we repeated the 
250000, 500000 and 2 million particle simulations, but with $\alpha_{\rm SPH} = 0.1$ and $\beta_{\rm SPH} = 2$.  
Figure \ref{fig:500k_comparison} shows the final state for two 500000 particle simulations, both of 
which used $\beta = 8$.  In the top panel, $\beta_{\rm SPH} = 0.2$, while in the bottom panel $\beta_{\rm SPH} = 2$.
Although they are similar, there are clear differences.  Using $\beta_{\rm SPH} = 2$ has removed some of the
noise present in the outer parts of the $\beta_{\rm SPH} = 0.2$ simulation.  However, although
there is coherent spiral structure in the inner parts of the $\beta_{\rm SPH} = 0.2$ simulation, it
is not present in the $\beta_{\rm SPH} = 2$ simulation.  Additionally, the inner hole is larger 
when $\beta_{\rm SPH} = 2$ than when $\beta_{\rm SPH} = 0.2$.  This is consistent with shocks being
more smeared out when a larger 
$\beta_{\rm SPH}$ value is used and is consistent with the larger $\beta_{\rm SPH}$ producing a larger
artificial viscosity (and hence clearing out more of the inner disc).

\begin{table}
\centering
\begin{minipage}{140mm}
  \caption{List of the simulations using smoothed cooling with $\beta_{\rm SPH} = 2$.} \label{tab:sims2}
  \begin{tabular}{c || ccc}
  \hline
  \hline
   Simulation & No. of particles &  $\beta_{\rm cool}$ & Fragment?  \\
 \hline
  1 & 250000 & 4 & Yes  \\
  2 & 250000 & 5 & Yes \\
  3 & 250000 & 6 & No  \\
  4 & 250000 & 7 & No  \\
  6 & 500000 & 5 & Yes  \\
  7 & 500000 & 6 & No \\
  8 & 500000 & 7 & No  \\
  11 & 2000000 & 7 & Yes \\
  12 & 2000000 & 8 & No \\
 \hline
  \hline
\end{tabular}
\end{minipage}
\end{table}

\begin{figure}
\begin{center}
\includegraphics[scale=0.5]{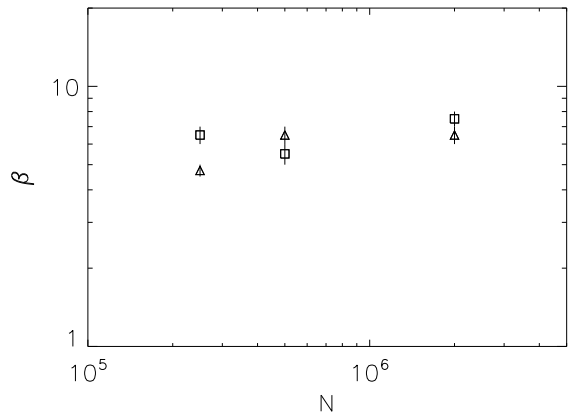}
\caption{Comparison of the fragmentation boundary obtained using smoothed cooling with $\beta_{\rm SPH} = 2$ (squares) and with 
$\beta_{\rm SPH} = 0.2$ (triangles).  The symbols are located at the average of the 
maximum $\beta_{\rm cool}$ for which fragmentation occured and the minimum for which it didn't, while the lines 
indicates the range between these two values.}
\label{fig:fragbound_bSPH2}
\end{center}
\end{figure}

Table \ref{tab:sims2} shows the results of the smoothed cooling simulations using $\beta_{\rm SPH} = 2$. Fig. \ref{fig:fragbound_bSPH2}
compares the results using $\beta_{SPH} = 0.2$ (triangles) with those obtained using $\beta_{\rm SPH} = 2$ (squares).
The symbols are located at the average of the maximum $\beta_{\rm cool}$ for which fragmentation occured and the 
minimum for which it didn't.  The short lines indicate the range between these two values.  Although there is a difference,
the results are very similar.  There is no indication that $\beta_{\rm crit}$ increases by 50\% when $\beta_{\rm SPH} = 2$,
compared to that obtained when $\beta_{\rm SPH} = 0.2$.  Given that previous analysis \citep{meru12} indicates
that $\beta_{\rm SPH} = 2$ should produce more artificial dissipation than $\beta_{\rm SPH} = 0.2$, it is a little surprising
that $\beta_{\rm crit}$ isn't smaller when using $\beta_{\rm SPH} = 2$ than when using $\beta_{\rm SPH} = 0.2$.  
It is possible that using $\beta_{\rm SPH} = 2$ does reduce some of the random noise associated with SPH, but it
is also clear (from Fig. \ref{fig:500k_comparison}) that it also changes the shock structure in the disc.  Maybe it
is not that surprising that the results differ slightly.  These results are, however, not consistent with 
the suggestion \citep{meru12} that - even for large particle numbers - artificial dissipation 
provides $\sim 30$\% of the heating when $\beta_{\rm SPH} = 0.2$. 

\section{FARGO}
\label{sec:fargo}
In addition to considering how self-gravitating discs evolve in SPH simulations, \citet{meru12} have extended this to consider grid-based simulations of self-gravitation accretion discs.  They use the {\sc fargo} code and show too that the fragmentation boundary does not converge as resolution increases. In this work, they also vary the artificial viscosity parameter so as to maximise the cooling time at which fragmentation occurs. There are, however, some issues with how they have implemented these {\sc fargo} simulations.

As with SPH, grid-based methods such as {\sc fargo} \citep{masset00} require a form of artificial viscosity to handle shocks. This is a direct consequence of Godunov's theorem \citep{godunov54}: any numerical scheme that is better than first-order accurate will introduce unphysical oscillations in the flow near shocks. Since almost all numerical methods for gas dynamics aim for at least second-order accuracy, a special recipe is needed around discontinuities in the flow. Finite-difference methods like {\sc zeus} \citep{stone92} and {\sc fargo} employ a van-Neumann-Richtmeyer type of artificial viscous pressure, which, when considering an axisymmetric disc, takes the form:
\begin{equation}
P_\mathrm{av}=\left\{
\begin{array}{ll}
Q_\mathrm{av}^2 \Sigma \left(\frac{\partial v_r}{\partial r}\right)^2 & \mathrm{if}~\partial v_r/\partial r<0,\\
0 & \mathrm{otherwise,}
\end{array}\right.
\label{eqartvisc}
\end{equation}
where $\Sigma$ denotes the surface density, $v_r$ the radial velocity and $Q_\mathrm{av}$ is the artificial viscosity parameter, which has dimensions of length. The reduced artificial viscosity parameter $q=Q_\mathrm{av}/\Delta r$, where $\Delta r$ is the grid spacing, determines over how many grid points shocks will be smeared out. In this view, of course, only values of $q$ larger than unity make sense, and the standard value in {\sc fargo} is $q=1.41$. Choosing a nonlinear viscous pressure as artificial viscosity results in the correct entropy jump across shocks and the correct shock propagation velocity \citep{neumann50}. 

The form of artificial viscosity given in equation (\ref{eqartvisc}) has two important properties: it acts only when the flow is compressed\footnote{Therefore, unlike as suggested in \cite{meru12}, artificial viscosity does \emph{not} act on the Keplerian shear. This is still true if a tensor form of the artificial viscosity is used, as long as the off-diagonal terms of the stress tensor are dropped \citep{stone92}.}, and it acts, for $q$ of order unity, only on length scales of the order of the grid scale. This latter property implies that if the value of $q$ makes a difference in the outcome of a simulation, the flow \emph{must} be under resolved. 

We illustrate the effect of artificial viscosity on two one-dimensional (axisymmetric) problems below, one linear and one non-linear. 

\subsection{Linear problem}

\begin{figure}
\begin{center}
\resizebox{\hsize}{!}{\includegraphics[]{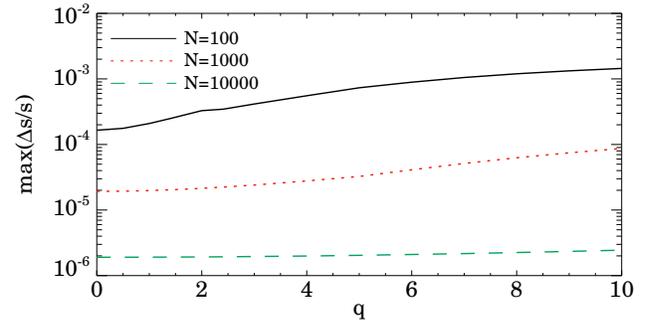}}
\caption{Maximum fractional change in entropy in the linear test problem as a function of the artificial viscosity parameter $q$, for three different grid sizes.}
\label{fig:fargolinear}
\end{center}
\end{figure}

For the first problem, we take an equilibrium inviscid Keplerian disc, extending from $r=0.5$ to $r=1.5$, with $H/r=0.05$ at $r=1$, and add a radial velocity perturbation
\begin{equation}
v_r = 0.01\exp\left(-\frac{(r-1)^2}{0.001}\right).
\end{equation}
Note that this is a velocity perturbation equal to $20$ \% of the sound speed. Therefore, no shocks form in this problem, which means that entropy should be materially conserved. For simplicity, we take the surface density to be constant initially, and choose the pressure so that the initial state has constant entropy ($P\propto\Sigma^\gamma$, where we take the ratio of specific heats $\gamma=1.4$), which means that in an ideal world, the quantity $s=P/\Sigma^\gamma$ should remain constant. No numerical method is ideal, of course, and there are two sources of changes in $s$: one is due to the finite size of the grid cells, which, unless a special entropy-conserving integration scheme is adopted, will lead to spurious changes in $s$, and the other is artificial viscosity, which directly changes the entropy through the viscous heating term. 

The results after integrating to $t=2.2$ are displayed in figure \ref{fig:fargolinear} for three different grid sizes $N$, with corresponding resolutions $\Delta r=1/N$. The initial Gaussian pulse in velocity is resolved by $\sim 3$ cells for $N=100$, $\sim 300$ cells for $N=1000$ and $\sim 3000$ cells for $N=10000$. As the artificial viscosity is increased, the maximum change in $s$ increases due to viscous heating, as expected. 

The increase in entropy at $q=0$ is due to the grid only. At the lowest resolution, the effects of the grid and the artificial viscosity are of similar magnitude. {\sc fargo}, like most grid-based methods, is second order accurate. This means that after a fixed number of time steps, errors should decrease as $\Delta r^2$ as the grid is refined. Since the number of time steps required to reach $t=2.2$ is proportional to $N$, we expect the errors due to the grid at $t=2.2$ to decrease as $\Delta r$, which is exactly what is observed in figure \ref{fig:fargolinear}.     

As the resolution is increased, the differences between runs with $q=0$ and $q\sim 1$ decrease. For all resolutions except $N=100$, taking $q\sim 1$ makes no difference compared to $q=0$. In other words, for $N=1000$ and $N=10000$, heating due to artificial viscosity is completely negligible for $q\sim 1$. Only for $N=100$ does artificial viscosity make a difference, but this is to be expected, since the extent of the initial pulse comparable to the grid scale, which means it will feel the artificial viscosity. Therefore, as expected, artificial viscosity plays no role in heating a smooth flow, unless it is not resolved.   

\subsection{Nonlinear problem}

\begin{figure}
\begin{center}
\resizebox{\hsize}{!}{\includegraphics[]{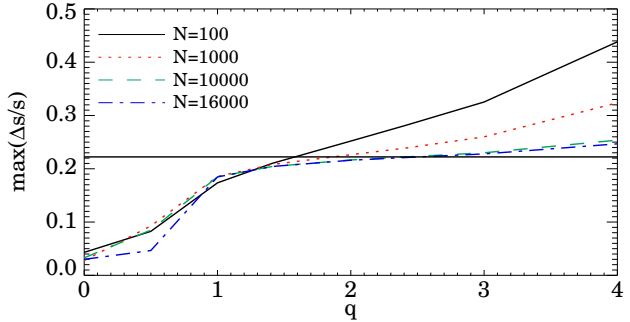}}
\caption{Maximum fractional change in entropy for the nonlinear test problem as a function of the artificial viscosity parameter $q$, for four different grid sizes. The horizontal line shows the solution obtained with a Riemann solver at $N=10000$. }
\label{fig:fargononlinearq}
\end{center}
\end{figure}

\begin{figure}
\begin{center}
\resizebox{\hsize}{!}{\includegraphics[]{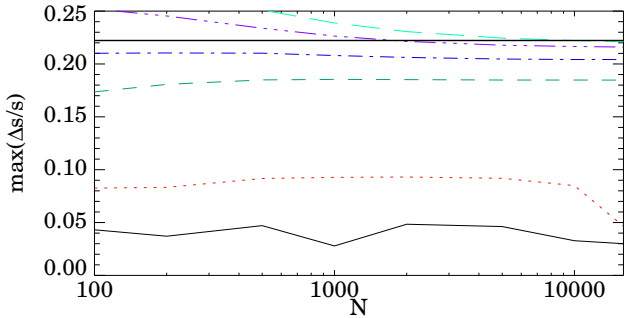}}
\caption{Maximum fractional change in entropy for the nonlinear test problem as a function of the grid size $N$. From bottom to top, $q=0,0.5,1.0,1.41,2.0,2.41$. The horizontal line shows the solution obtained with a Riemann solver at $N=10000$. }
\label{fig:fargononlinearN}
\end{center}
\end{figure}

As a second test problem, we set up a nonlinear wave characterised by initial conditions
\begin{equation}
\Sigma=\left\{
\begin{array}{ll}
1 & \mathrm{if}~r<1,\\
0.1 & \mathrm{otherwise,}
\end{array}\right.
\label{eqshock}
\end{equation}
in the same inviscid equilibrium Keplerian disc as above. The initial pressure is set again such that the initial state is isentropic. The solution develops a shock that leads to an increase in entropy. The correct increase in $s$ was measured from a simulation using a Riemann solver \citep{paardekooper06} at $N=10000$. 

The results obtained with {\sc fargo} are displayed in figure \ref{fig:fargononlinearq}. It is immediately clear that simulations with $q\leq 1$ strongly underestimate the change in entropy. This is again not surprising, since artificial viscosity is \emph{necessary} in this case because of the presence of an entropy-generating shock. Similar to SPH, choosing the artificial viscosity parameter too high leads to unphysical, artificial heating: choosing $q>2$ leads to too much viscous heating for $N=100$, but at high enough resolution a plateau emerges giving roughly the same amount of entropy generation independent of $q$.  

In figure \ref{fig:fargononlinearN} we look at the same problem but now as a function of resolution. Simulations with $q<2$ systematically underestimate the entropy production, independent of resolution. The situation is most severe for $q=0$ and $q=0.5$, which do not even show convincing signs of convergence with resolution. Simulations with $q\geq 1$ do seem to converge, but to a level that depends on $q$. For the largest values of $q$, the entropy increase converges to a value very close to the correct one, but at the price of overestimating the entropy increase at low resolution. The standard value $q=1.41$ seems to be a good compromise. 

\subsection{Implications for self-gravitating disc simulations}
It is not completely straightforward to translate the above results to two dimensions. In this case, an extra source of error comes from dimensional splitting, the effects of which are not entirely clear especially if the {\sc fargo} algorithm is used \citep{masset00}. Moreover, shocks are no longer necessarily aligned with the grid, which in all likeliness changes the dissipation properties of the grid. However, a few general statements can be made. 

The 'optimum' value of $q$ is larger than unity. For example, $q=1.41$ gives artificial viscosity that leads to a good estimate of entropy production in shocks (figure \ref{fig:fargononlinearq}), while its effect on \emph{resolved} smooth flow is negligible (figure \ref{fig:fargolinear}). Values of $q<1$ do not give the correct amount of shock heating, and can therefore not be expected to give physical results. Note that the 'optimum' value chosen by \cite{meru12}, $q=0.5$, underestimates the entropy increase in a shock by a factor of 3 (see figure \ref{fig:fargononlinearN}). This will have serious consequences for the simulation results if heating is due to shock dissipation, which is the case for self-gravitating discs.  

If the amount of artificial viscosity makes a difference in the results, the flow is under resolved. This can be due to shocks, in which case there the flow can in fact not be resolved, or due to unresolved smooth flow. In the latter case, reducing the amount of artificial viscosity will in general not lead to a much better solution, since errors due to the finite size of the grid cells are likely to be as large as the error introduced by artificial viscosity, precisely because  the flow is unresolved (figure \ref{fig:fargolinear}). Moreover, reducing the amount of artificial viscosity can only be safely done when it is absolutely certain that no shocks are present in the problem, which, for self-gravitating discs, we know is not the case. 

Keeping the artificial parameter fixed at $q=1.41$, there are still several avenues for investigating the problem of convergence in grid-based simulations. A direct comparison between 2D global and local simulations \citep{gammie01}, which do not appear to show convergence \citep{paardekooper12}, is definitely warranted. Care must be taken in global simulations to avoid initial transients \citep{paardekooper11}. It may be that the lack of convergence in grid-based simulations is the result of the two-dimensional approximation, even though smoothing of the gravitational potential does not seem to make much of a difference \citep{paardekooper12}. It may also be that the inviscid problem is ill-posed, and that a finite amount of (Navier-Stokes) viscosity is needed to reach convergence.   

\section{Importance of artificial viscosity in SPH and {\sc fargo}}
Artificial viscosity is a necessary feature in numerical simulations in order to make the code behave in a reasonable way at the smallest resolvable scale, which is the scale of the grid in grid-based codes, and the smoothing length in SPH. Without artificial viscosity, particle interpenetration would make the outcome of any SPH simulation useless. The use of a grid introduces its own associated 'viscosity', which acts on the smallest resolvable scale (the grid scale) and may or may not behave like a real viscosity. This is why simulations of turbulence often employ an additional physical Navier-Stokes viscosity to make sure that energy dissipation on the smallest scales is well-behaved and physical (see e.g. Fromang, Papaloizou , Lesur, \& Heinemann 2007). In addition, both SPH and grid-based codes like {\sc fargo} need artificial viscosity to handle shocks correctly. 

It should be clear from the discussion in sections \ref{sec:results} and \ref{sec:fargo} that care should be taken when trying to adjust the amount of artificial viscosity for a particular problem, both in grid-based codes and SPH. When the level of artificial viscosity makes a difference in the simulation outcome, this means that features close to the smallest resolvable scale (either the grid scale or the smoothing length) play an important role. Artificial viscosity is a way of converting bulk motion on these small scales into heat. This can be unphysical, if the underlying flow is smooth, or physical, in the case of shocks. Of course, one would always like to reduce the amount of artificial viscosity as much as possible. However, reducing the amount of artificial viscosity below the level required to handle shocks correctly, e.g. taking $q<1$ in {\sc fargo}, can only be done if it is known that no shocks will occur in the problem. And even then, the results in section \ref{sec:fargo} indicate that for smooth flow dissipation is dominated by the grid rather than artificial viscosity, so that changing $q$ will not make a difference in the simulation outcome. 

The only way to do a better job for smooth flow in a grid-based simulation is to increase the resolution. In SPH, there are more artificial viscosity parameters to play with, and it may be slightly less clear to what extent the choices for $\alpha_{\rm SPH}$ and $\beta_{\rm SPH}$ are free. Since there is no dissipation due to a grid, it is likely that the level of artificial viscosity matters more compared to grid-based codes. Typically SPH simulations use $\beta_{\rm SPH} = 2 \alpha_{\rm SPH}$.  This is so that the linear term dominates when particles are converging slowly, and the quadratic term dominates when the particles are converging fast enough that shocks are likely to form.  Using $\beta_{\rm SPH} = 20 \alpha_{\rm SPH}$, as suggested by \citep{meru12}, effectively means that the quadratic term is likely to always dominate and, hence, is likely to change the properties of the simulation itself.

In the case where shocks are present, the situation is more complicated, since the numerical method will always smear shocks over a few grid cells, for a grid-based code, or a few smoothing lengths for SPH. Increasing the resolution therefore keeps reducing the size of shocks, which means that ever smaller scales are present in the problem. These smallest scales can interact in non-trivial ways with larger scale structures (clumps, waves), and it may not be immediately clear whether convergence can be reached and at what resolution.

\section{Discussion and Conclusions}
Recently, \citet{meru11} have suggested that, in three-dimensional SPH simulations of self-gravitating accretion disc with
$\beta$-cooling, the cooling time ($\beta_{\rm cool}$) at which fragmentation occurs does not converge as resolution 
is increased.  They've extended this work \citep{meru12} to suggest that with typical artificial viscosity parameters, the
fragmentation boundary is converging towards a critical cooling time of $\beta_{\rm crit} = 17.4$.  They go on to argue, however, that adjusting the 
artificial viscosity parameters (so as to maximise the critical cooling time, $\beta_{\rm crit}$) suggests that, with an appropriate
choice of the artificial viscosity parameters, the simulations actually converge towards $\beta_{\rm crit} = 29.2$.  

\citet{meru12} then continue this by considering the evolution of self-gravitating disc using the grid-based code {\sc fargo}.  
Here they also vary the artificial visocsity parameter so as to maximise the critical cooling time and show that these simulations
also don't converge as resolution increases.

It has been suggested \citep{rice12} that the non-convergence seen in \citet{meru11} was simply a consequence of the manner in which
cooling was implemented.  We've extended the work of \citet{rice12} here to show that by implementing what they
call smoothed cooling, fragmentation requires $\beta_{\rm cool} < 8$ for all resolutions considered (from 250000 particles to
10 million particles).  This is more consistent with other work \citep{gammie01} and also makes physical sense given that
the Jeans mass is well resolved in most of the disc for simulations with 500000 particles or more, and that artificial viscosity 
should be providing less than 10\% of the heating in such simulations, so shouldn't be significantly influencing the fragmentation
boundary.  Furthermore, if fragmentation can occur for $\beta_{\rm cool} \sim 30$ this suggests that a clump can contract and become
bound even though the timescale over which it is losing energy is significantly greater than the orbital period, which is likely
to determine the timescale over which we'd expact the clump to heat.  

We also consider how the alternative artifical viscosity values suggested by \citet{meru12} influence the results when using
smoothed cooling.  We find that the results are consistent with those obtained using the original artificial viscosity parameters.  
Rather than increasing the critical cooling time, $\beta_{\rm crit}$, by 
50\% \citep{meru12}, fragmentation still requires $\beta_{\rm cool} < 8$, for all particle numbers considerd. 
If this change to the artificial viscosity parameters was reducing
the level of artificial heating (as suggested by \citet{meru12}) then we'd expect the results to be independent of the
implementation of the cooling.  That they aren't suggests that changing these parameters is influencing the 
simulations in some numerical way, rather than simply changing the level of artificial heating - especially as the expectation
is that the changes made by \citet{meru12} should have increased, rather than reduced, the level of artificial heating.  
This is also consistent with the significant difference 
between the spiral shock structure in the disc with $\beta_{\rm SPH} = 2$ when compared to simulations with $\beta_{\rm SPH} = 0.2$.

What was quite attractive about the \citet{meru12} work was that they obtained very similar results when using the grid-based
{\sc fargo} code.  However, here, they also varied the artificial viscosity parameter so as to minimise the level artificial viscosity. 
As discussed earlier, however, the artificial viscosity in {\sc fargo} only acts on converging flows and so should not produce
any artificial (non-shock related) heating.  Additionally, in order to produce good estimates of entropy production at shocks
requires that, as discussed earlier, the optimal value for the artificial viscosity in {\sc fargo} be larger than unity. 
\citet{meru12} claim that the optimal value is $q = 0.5$, much smaller than would be regarded as suitable for such a simulation.
Admittedly, even their simulations with $q = 1.41$ did not show signs of convergence, but this could be related to unsuitable
initial conditions \citep{paardekooper11} or the possibility of stochasticity \citep{paardekooper12} and should be investigated
further.

Essentially, the SPH simulations presented here show that if one implements the cooling so as to remove the unphysical discontinuity
at the contact discontinuity behind shocks \citep{price08,price12} we appear to get convergence as the resolution increases,
and the fragmentation boundary that we determine is consistent with earlier work (fragmentation occuring for cooling times between
$\beta_{\rm cool} = 6$ and $\beta_{\rm cool} = 8$ for $\gamma = 5/3$). Although, we haven't investigated \citet{meru12}'s 
{\sc fargo} results in as much detail, it seems clear that what they regard as the optimal value for the artificial viscosity
parameter ($q = 0.5$) is well below what would be regarded as acceptable for such simulations.  This might suggest that 
their {\sc fargo} results suffer from additional numerical issues.  With the exception of the possibility of stochastic
fragmentation \citep{paardekooper12} we therefore conclude that there is no real evidence that fragmentation can occur
in self-gravitating discs with long cooling times and that the likely fragmentation boundary is similar to that
suggested by earlier work \citep{gammie01,rice05}.  Consequently, this implies that - as suggested by earlier 
studies \citep{rafikov05, stamatellos08, clarke09, rice09} - gas giant planet formation via disc fragmentation 
is unlikely in the inner regions ($r < 50$ au) of protostellar discs.

\section*{Acknowledgements}

\noindent 
All simulations presented in this work were carried out using high performance computing funded by the Scottish Universities Physics
Alliance (SUPA). WKMR and DHF gratefully acknowledge support from STFC grant ST/J001422/1. PJA ackowledges support 
from NASA's Astrophysics Theory and Origins of Solar Systems programs through grants NNX11AE12G and NNX13AI58G.

\label{lastpage}

\end{document}